\documentclass[12pt]{iopart}
\usepackage{mdframed}
\usepackage{graphicx}
\usepackage{url}
\usepackage{array}
\begin{document}

\title{Reverse resonant nonspecular reflection controlled by intensity of light}

\author{Sergiy Prosvirnin$^{1, 2}$, Vyacheslav Khardikov$^{1, 2}$,  Vladimir Yachin$^{1}$, Vadym Plakhtii$^{2}$, and Nataliia Sydorchuk$^{1}$}

\address{$^1$ Institute of Radio Astronomy, Kharkiv 61002, Ukraine }
\address{$^2$ Karazin Kharkiv National University, Kharkiv 61022, Ukraine}
\ead{prosvirn@rian.kharkov.ua}

\begin{abstract}
Nonlinear optical features of a plane structure consisted of silicon disks placed periodically on a silver substrate have been studied in the Littrow reflection scenario. The structure manifests a bistable resonant reflective ability.  Values of both specular reflectance and reverse one can be tuned by varying an intensity of incident light.
\end{abstract}

\section{Introduction}

A flat metallic mirror is a simplest optical device  being used since antiquity \cite{ancient_mirrors-2006}. A remarkable feature of a mirror is exhibited by extremely useful capability of changing sharply the direction of light rays and returning light back at normal incidence over a wide spectral bandwidth. Today photonics requires mirrors with specific reflection selectivity both in electromagnetic field intensity and phase depending on a light wavelength. Spatial variation  of the field phase along the reflection surface is specially adapted in order to manipulate, redirect and concentrate light intensity.  

Metasurfaces open prospects to create specific reflectors with extraordinary properties \cite{metasurface-2016}. The  wavelength-selective properties of mirrors are necessary to design precise sensors \cite{absorber_for_sensor_2015, multispectral_sensing_2015, sydorchuk-2017}, in particular, biological material sensors \cite{sensors-2017}.  Another aspect of the application of resonant mirrors involves placement of a certain quantum system which can have different energy states in an intense field near the reflecting surface. In this case, the challenge is to obtain the maximum radiation intensity or extra absorption. In the light of these problems, along with classical metasurfaces, resonant arrays radiated only one or a few diffracted orders besides main one look attractive for practical applications due to their promising features \cite{Collin_2014, Zhu:15}. 

Modern optics technologies give the opportunity to produce resonant non-specular and selectively reflecting metasurfaces. A reflecting metasurface is usually a planar double periodic subwavelength patterned metal-dielectric or all-dielectric layer placed on a metal substrate \cite{meta_mirrors-2014}. The thickness of a metasurface is usually very small in comparison with an electromagnetic  wavelength in free space. However, the periodic structured surface provides conditions for the excitation of different types of resonances. A spectacular metasurface response manifests in resonance reflection and absorption \cite{meng-2017}, in electromagnetic radiation enhancement by using a laser medium \cite{yablonovitch-2015}, in exotic electromagnetic field boundary values, which may be the same as on a surface of an artificial magnetic wall \cite{4666749}, and in the confinement of the intensive electromagnetic field inside the structure \cite{all-dielectric-2016}.

Among numbers of important electromagnetic properties, the ability of flat periodic structures to reflect an incident wave in a preferable non-specular direction is extremely attractive for applications in photonics \cite{meta_mirrors-2014, metasurface-2016}. The first researches of a subject have long history and initially were related to single-periodic reflective gratings for suppressing unwanted reflection  of microwave radiation \cite{shestopalov-1973, 1141640, Masalov-1980}. In laser technology of optical electronics, inclined echelette grating mirrors were used to reflect light in the direction strictly opposite to the direction of arrival of the incident wave. The corresponding scenario of diffraction by a structure periodic in one direction  is known as the Littrow scheme \cite{hard-1970, Lotem-1994} or an autocollimation regime. Another practically important diffraction scenario is when the obliquely incident wave is reflected in the direction normal to the metasurface.  In particular, it is of interest for the design of quasi-optical power pulse  compression devices \cite{Gribovsky_2014}.

As it is known, in a particular diffraction scenario, the propagation directions of diffraction orders depend only on a geometric shape and sizes of a unit cell of the periodic metasurface in comparison with the wavelength and on the direction of the incident electromagnetic wave.  On the other hand, power distribution between diffraction orders is determined only by resonance properties of meta-atoms that make up the periodic structure and their mutual coupling. Therefore, a change in the resonance properties of the meta-atoms opens up the possibility of switching the structure reflectivity between states of specular and non-specular reflection.

For changing the properties of meta-atoms, one of opportunities is to use materials in which polarizability depends on intensity of the applied electromagnetic field \cite{Litchinitser_2015}. In this case, switching between the states of specular and non-specular reflection of light is achieved by changing the intensity of the incident wave. In particular, the optical Kerr effect is defined as an intensity dependent refractive index $n=n_0+n_2I$ where $I$ is the intensity of the optical field, $n_0$  is the  linear refractive index in the low light intensity regime and $n_2$ is the nonlinear refractive index. 

The optical Kerr effect is a third-order effect, which can manifest optical bistability of the structure response. This phenomenon can be applied to design devices as a toggle switch. Third order nonlinear optical properties is inherent to many different media (both centrosymmetric and noncentrosymmetric), in particular silicon that is most using medium of photonics \cite{boyd-2008}.

In this paper, for the first time, we claim a theoretical demonstration of the controllable features of the uncommon resonant regime of {\it {the reverse reflection}} realised with the patterned silicon-on-metal structure. The regime is associated with excitation of one additional diffraction order besides the main partial wave.  Ratio of the non-specular to specular reflectance values can be controlled by a variation of the incident wave intensity. To provide the non-specular reflection for any incident wave direction we have proposed an array of silicon disks with the excitation of the Mie-resonance that has a specific symmetry of  the field distribution \cite{prosvirnin2021nonspecular} used in antennas for mobil communication. 

In our research we have assumed electromagnetic fields of combination frequencies  negligibly small and considered the field as monochromatic. A basis for such approach is the use of resonant elements on the metasurface. The research was fulfilled with a finite element method and full-wave numerical simulation. 

\section{Problem statement}
\label{problem statement}

\begin{figure}[htbp]
\centering
\includegraphics[width=4 in]{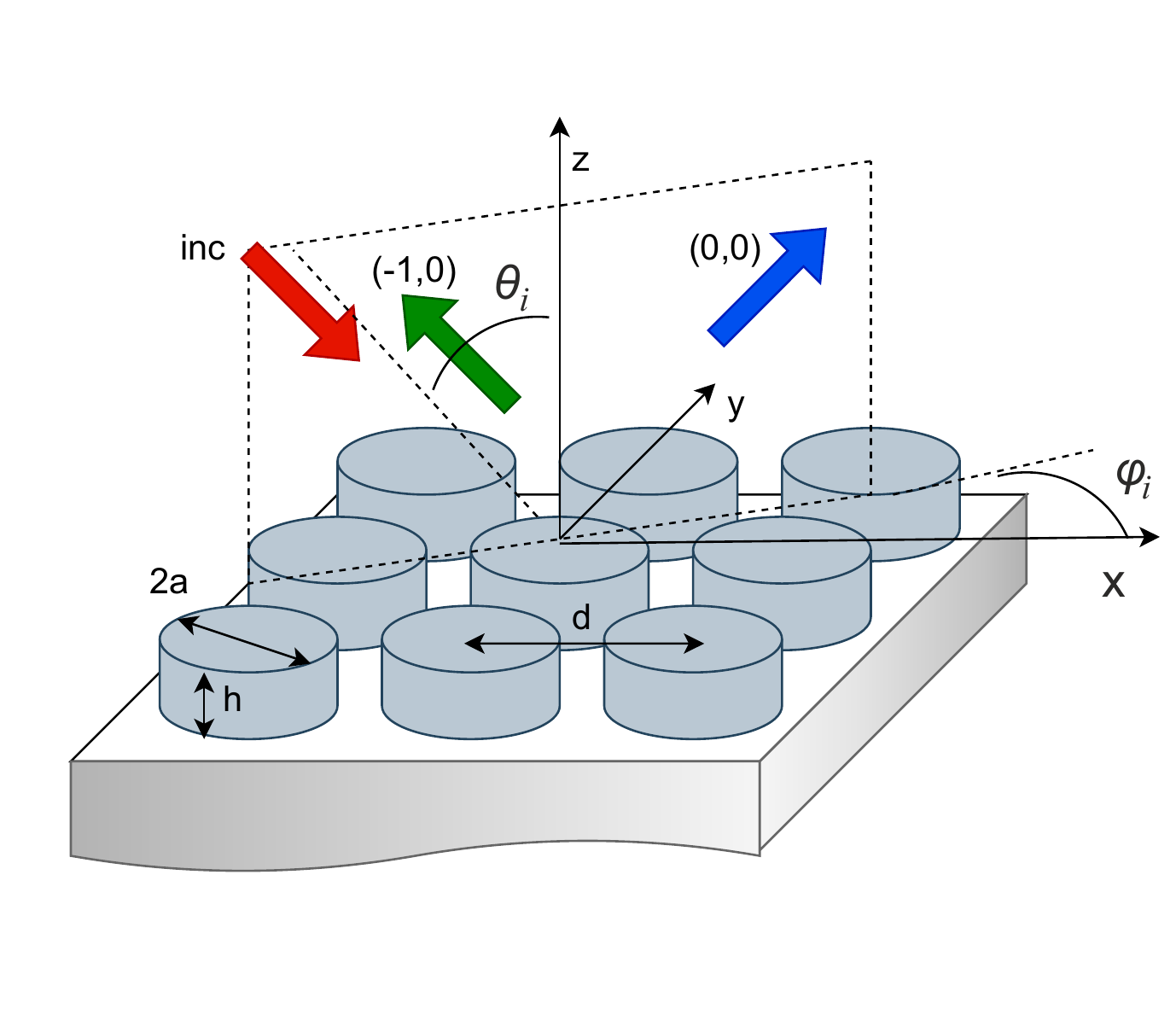}
\caption{Schematic of the doubly-periodic planar reflective structure under study. The incident ray (red arrow) and also specular and reversely reflected ones (blue and green arrows respectively) are placed in the plane of incidence which is shown by a dashed parallelogram. }
\label{fig_1}
\end{figure}

Let us study reflection of the plane electromagnetic wave
\begin{equation}
 \label{Ei}
  {\bf{E}}^{i}={\bf e}^{i} E \exp(-i{\bf k}^{i} {\bf r}) 
\end{equation}
by a doubly-periodic reflecting structure placed in free space parallel to the $xy$-plane (see Fig. \ref{fig_1}). In expression (\ref{Ei}), the vector ${\bf e}^{i}$ is an unit polarization vector, $E$ is a field strength of the incident wave and 
$${\bf k}^{i} = {\bf e}_x k \sin\theta_i \cos\varphi_i + {\bf e}_y k \sin\theta_i \sin\varphi_i - {\bf e}_z k \cos\theta_i $$ 
is a wave vector, where $\varphi_i$ and $\theta_i$ are the azimuth and polar angles of the incidence direction respectively, $k=\omega/c=2\pi/\lambda$, vectors ${\bf e}_x$, ${\bf e}_y$ and ${\bf e}_z$ are unit basis vectors along the axes $x$, $y$ and $z$.  

In the region $z>0$, the field is a superposition of the incident wave field and fields of partial diffracted waves propagating away from the structure and along its surface:
\begin{equation}\label{E1}
  {\bf{E}}= 
    {\bf{E}}^{i} + \sum_{m,n=-\infty}^\infty
    {\bf{a}}_{mn}
    \exp( - i{\bf{k}}_{mn} {\bf r}), \qquad z>0,
\end{equation}
where ${\bf{a}}_{mn}$ and ${\bf{k}}_{mn} =
{\bf{g}}_{mn} + {\bf e}_z\gamma_{mn}$ 
are vector amplitudes and wave
vectors of partial waves of the reflected
field respectively, 
\begin{equation}
\label{g_mn}
    {\bf g}_{mn} = {\bf e}_x (k_x^i + \frac{2 \pi m}{d}) +
    {\bf e}_y ( k_y^i + \frac{2 \pi n}{d})
\end{equation}
is projection of the wave vector of Floquet harmonic on the $xOy$ plane, 
$$
    \gamma_{m n} = \sqrt{k^2 - g_{mn}^2}, \quad \mbox{Re}
\gamma_{mn}\geq 0, \quad \mbox{Im}
\gamma_{mn}\leq 0
$$
is the component of the wave vector along the $Oz$-axis.

Each spatial harmonic having nonzero real component  $\gamma_{m n}$  of the wave vector corresponds to one of the plane waves, which transport energy from the array plane to the free space. The spatial harmonic
with subscripts $m = 0$  and $n = 0$ is a plane wave propagated
along a direction of specular reflection of the incident wave.  Thus, propagation directions of spatial harmonics or diffraction orders of the reflected field are defined by pitches of the array, a wavelength and a direction of the incident wave. However, an intensity distribution of spatial harmonic spectrum depends on the scattering properties of the structure meta-atoms.

For the sake of simplicity and shortness we have further considered the case $\varphi_i =0$  
The incident wave is supposed to have TE-polarization i.e. ${\bf e}^i={\bf e}_y$.

A normalized frequency defined as $\kappa=d/\lambda$ is convenient  for using. An equation $\gamma_{mn}(\kappa)=0$ is the equation on cutoff frequencies $\kappa_{mn}$ for the diffraction orders. If  a normalized frequency is lower then $\kappa_{-10}=(1+\sin\theta_i)^{-1}$ the only specular ray propagates away from the metasurface. A cutoff frequency of the diffraction order with $m=-2$ and $n=0$ is $\kappa_{-20}=2(1+\sin\theta_i)^{-1}$.  In the frequency band $(\kappa_{-10}, \kappa_{-20})$ a far-zone reflected field consists of two rays which are (0,0) and (-1,0) diffraction orders.

At the normalized frequency $\kappa=(2\sin\theta_i)^{-1}$, the diffraction order (-1,0) propagates strictly opposite to the incident wave propagation direction as can be easily seen \cite{Lotem-1994}.

When choosing resonant elements for the metasurface, our preference has been made for the silicon disks. Silicon was proposed as it is a most used dielectric material of photonics. The resonant features of a single dielectric disk located in free space and on a dielectric substrate are well studied  \cite{vandeGroep:13}. The plane arrays of disks were also studied in details in the case when their unit cell is smaller than a wavelength \cite{all-dielectric-2016}. Next, the attractive opportunities of using the electromagnetic features of dielectric disks placed on a metal substrate as an antenna elements of the mobile communication devices were discussed in \cite{dielectric-antenna-2011}. 

\section{Results of simulation and analysis}

The silicon disks of the array are assumed to be placed periodically on a silver plane thick  substrate (see Fig. \ref{fig_1}).  The array pitch $d$ is chosen equal to 750 nm. The disk radius is $a=130$ nm. The disk thickness is $h=130$ nm.  The linear and nonlinear refractive indexes of silicon are approximated by data presented in  \cite{Dinu_2003, Gholami_2011, Wang:13}. At frequencies close to the frequency of reverse reflection, they are $n_0 = 3.48$ and $n_2 = 10^{-18}$ m\textsuperscript{2}/W respectively.  The substrate thickness is assumed so large that the light field does not penetrate via the substrate. The refractive index of silver are referenced from \cite{werner-2009}. The angle of incidence is assumed $\theta_i=35$ degrees.

Such choice of the structure parameters gives us an opportunity to research a resonant diffraction efficiency of the specular reflected wave and (-1,0) diffraction order propagating opposite to the incident wave.

\begin{figure}[htbp]
\centering
\includegraphics[width=5 in]{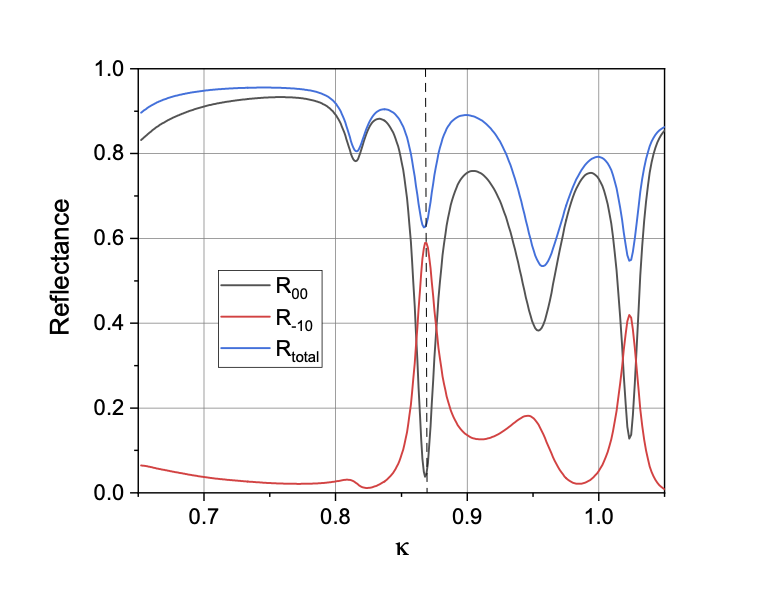}
\caption{Reflectances versus the normalized frequency in linear regime. The approximate normalized frequency value of  0.872 corresponding to the auto-collimation diffraction scenario has been marked by a dashed vertical line.}
\label{fig_2}
\end{figure}

In linear regime the designed reflect-array manifests the strong resonant response at the frequency close to the frequency of auto-collimation diffraction scenario by excitation of intensive (-1,0) diffraction order. In the Fig. \ref{fig_2}, a frequency dependance of reflectance that is a measure of how much optical power is reflected into the designated direction compared to the power incident onto the reflective item has been presented.

The designed structure manifests sharp resonant excitation of (-1,0)-diffraction order due to its adjustment to Mi-resonance of the silicon disk elements. The resonant field type is close to TE \textsubscript{01$\delta$} resonance of the dielectric cylinder of a finite length in free space. 

A value of resonant transformation of the specular ray intensity  to intensity of the reverse reflection ray depends on  dissipative properties of the reflect-array. In the base of a non-dissipative theoretical model of the reflecting structure, we have shown a complete transformation of the incident wave intensity into the intensity of the reverse wave (see \cite{9552327}). A power  dissipated by the array is defined by absorption value $A=(1-R_{total})$ (see Fig. \ref{fig_2}). 


\begin{figure}[htbp]
\centering
\includegraphics[width=5 in]{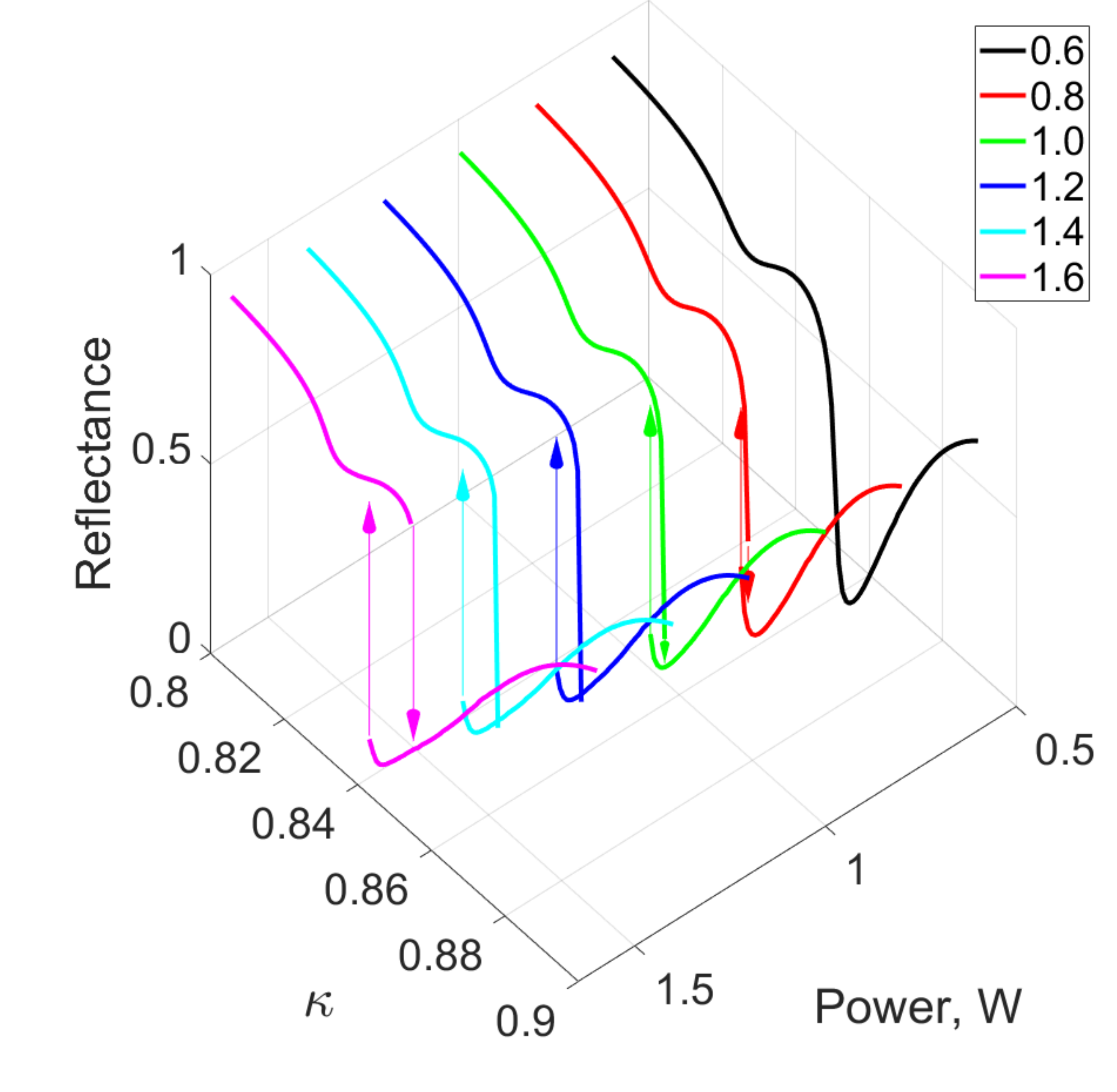}
\caption{Specular reflectance R\textsubscript{00} versus the normalized frequency and the incident wave power per unit cell of the reflect-array.}
\label{fig_3}
\end{figure}

\begin{figure}[htbp]
\centering
\includegraphics[width=5 in]{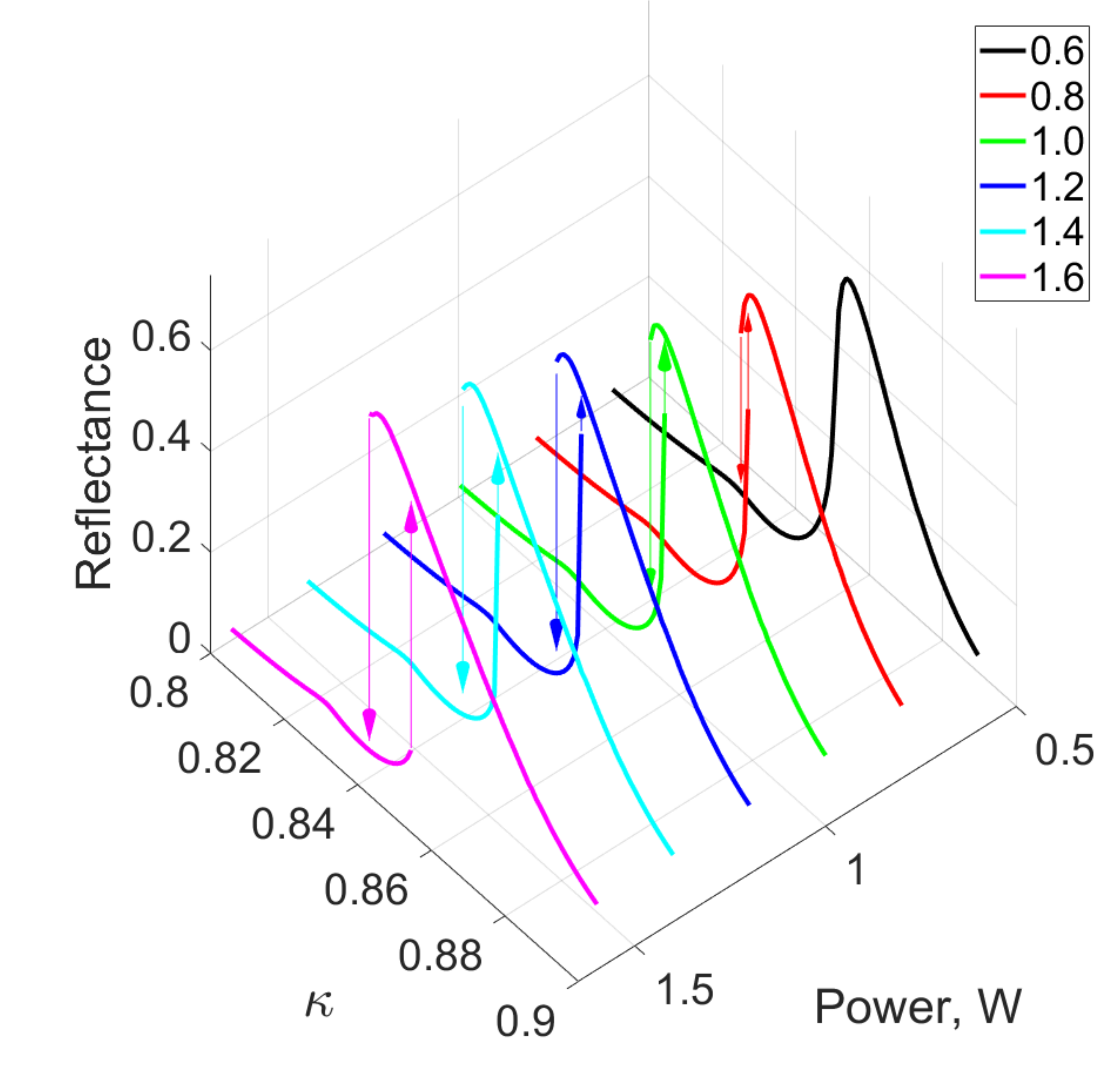}
\caption{Reverse reflectance R\textsubscript{-10} versus the normalized frequency and the incident wave power per unit cell of the reflect-array.}
\label{fig_4}
\end{figure}

Simulation results for the frequency dependences of the array reflectance are presented in figures \ref{fig_3} - \ref{fig_4} versus the power of the incident wave per unit cell of the array. We have observed a bistable reflection properties of the array if the power exceeds some value.  Our estimation of the power threshold of the reflection bistability is approximately 0.7 W.

Under bistable conditions the value of reflectance at each power level depends on whether the bistable state is reached by transition from low or from high frequencies. With an increase in the intensity of the incident wave, the frequency zone of bistability expands and shifts to lower frequencies.

Frequency dependences of reflectance are the hystereses loops. In the bistability frequency  band, the value of reflectance depends on the kind of way to reach  bistable state.

\section{Conclusion}

In summary, we have shown by a full wave simulation that a structure of identical silicon disks placed periodically on a silver substrate can be used as an effectively excited and adjusted resonant nonlinear dielectric-metal hybrid reflect-array in the Littrow non-specular diffraction scenario. It is found that the nonlinear response of the silicon disks can be employed for realising the bistable reverse reflection which is controllable by varying the intensity of the incident wave. The nonlinear tunability of silicon-on-silver structure is promising to extend the working area of classical subwavelength metamaterials providing new opportunities in applications with light-matter interactions.

\ack
Authors are grateful to the National Research Foundation of Ukraine (2020.02/0218) and the National Academy of Sciences of Ukraine for a support of this work.


\section*{References}


\bibliographystyle{unsrt}

\bibliography{ContrRevRefl}


\end{document}